\documentclass{article}
\usepackage{amscd,amsmath,amssymb,bbm}
\usepackage{amsfonts}
\newcommand{\p}[1]{(\ref{#1})}

\newcommand{\bD}{\overline{D}{}}
\newcommand{\bLam}{{\overline{\Lambda}}{}}

\newcommand{\bxi}{{\bar\xi}}
\newcommand{\blam}{{\bar\lambda}}

\newcommand{\brho}{{\bar\rho}}
\newcommand{\tb}{{\bar\theta}}
\newcommand{\halpha}{{\hat\alpha}}

\newcommand{\be}{\begin{equation}}
\newcommand{\ee}{\end{equation}}
\newcommand{\bea}{\begin{eqnarray}}
\newcommand{\eea}{\end{eqnarray}}
\newcommand{\ba}{\begin{array}}
\newcommand{\ea}{\end{array}}

\newcommand{\nn}{\nonumber}

\def\im{{\rm i}}

\def\={\ =\ }

\def\Nf{$\cal N${=\,}4~}

\begin{document}
\thispagestyle{empty}
\vspace{2cm}
\begin{flushright}
\end{flushright}\vspace{2cm}
\begin{center}
{\Large\bf N=4, d=1 Supersymmetric Hyper-K\"{a}hler Sigma Models \vspace{0.5cm} \\with Isospin Variables}
\end{center}
\vspace{1cm}

\begin{center}
{\Large
Stefano Bellucci$\,{}^{a}$, Sergey Krivonos$\,{}^{b}$  and
Anton Sutulin$\,{}^{b}$
}\\
\vspace{1.0cm}
${}^a$ {\it
INFN-Laboratori Nazionali di Frascati,
Via E. Fermi 40, 00044 Frascati, Italy} \vspace{0.2cm}

${}^b$
{\it Bogoliubov  Laboratory of Theoretical Physics,
JINR, 141980 Dubna, Russia}
\vspace{0.2cm}
\end{center}
\vspace{3cm}

\begin{abstract}
\noindent We provide a Lagrangian formulation of \Nf
supersymmetric mechanics describing the motion of an isospin
carrying particle on conformal to hyper-K\"{a}hler spaces in a
non-Abelian background gauge field. In two examples we discuss in
details, this background field is identified with the field of
BPST instantons in the flat and Taub-NUT spaces.
\end{abstract}

\newpage
\setcounter{page}{1}

\setcounter{equation}0
\section{Introduction}
The \Nf supersymmetric mechanics provide a nice framework for the
study of many  interesting features of higher dimensional
theories. At the same time, the existence of a variety of
off-shell \Nf irreducible linear supermultiplets in $d=1$
\cite{GR, Tolik, FG, FG1, ikl1} makes the situation in one
dimension even more interesting, and just this prompted us to
investigate such supersymmetric models themselves, without
reference to higher dimensional counterparts. Being a
supersymmetric invariant theory, the \Nf mechanics admits a
natural formulation in terms of superfields living in a standard
and/or in a harmonic superspace \cite{hss}, adapted to one
dimension \cite{IL}. In any case the preferable approach to
discuss supersymmetric mechanics is the Lagrangian one.

Being
quite useful, the Lagrangian approach has one subtle point, when
we try to describe the system in an arbitrary gauge background.
Indeed, when the bosonic metric is conformally flat and the gauge field ${\cal A}_\mu$ is Abelian, the corresponding (bosonic)
Lagrangian can be immediately
written as
\be\label{intro1}
L\sim f(x) {\dot x}_\mu{\dot x}_\mu+{\cal A}_\mu(x){\dot x}_\mu+\ldots.
\ee
This Lagrangian can be properly extended by fermionic fields to enjoy
\Nf supersymmetry. It admits a nice superfield formulation \cite{IL,Kon1}.

If instead the gauge field is non-Abelian (matrix-valued), then, in order to write
the corresponding Lagrangian, one has to introduce extra variables $(\omega^i,{\bar \omega}_i)$
in the fundamental representation of $SU(n)$ \cite{Kon2}.
For the simplest case of the $SU(2)$ group the corresponding (bosonic) Lagrangian reads
\be\label{intro2}
L\sim f(x) {\dot x}_\mu{\dot x}_\mu+{\cal A}^a_\mu I^a {\dot x}_\mu
+ \frac{\im}{2}\,\left( {\dot \omega}{}^i{\bar \omega}_i - \omega^i{\dot{\bar \omega}}_i\right),
\ee
where
\be\label{intro3}
I^a \sim i \omega^i \left( \sigma^a\right)_i^j \bar \omega_j\,.
\ee
Due to the presence of only first order time derivatives in a very specific kinetic term for
the isospin variables $(\omega^i,{\bar \omega}_i)$ the latter are ``semi-dynamical''
ones and become purely internal
degrees of freedom after quantization.

The Lagrangian of the discussed type  \p{intro2} can be also made \Nf supersymmetric by adding
the corresponding fermions \cite{Kon1,Kon2,FIL,brazil,FIL1,bk,KL1,sutulin1,Ikon,sutulin2}.
Moreover, it turns out that for constructing the off-shell supersymmetric Lagrangian one has to include
in the game some additional auxiliary fermions which form the off-shell \Nf supermultiplet with bosonic
isospin variables $(\omega^i,{\bar \omega}_i)$. Thus, in order to deal with \Nf supersymmetric mechanics
describing the motion of the particle in some non-Abelian gauge background fields, one has to consider
the coupled system of ``matter'' superfields ($x_\mu$ and corresponding fermions) and the auxiliary
superfield which contains isospin variables $(\omega^i,{\bar \omega}_i)$ together with auxiliary fermions.

There are different approaches to introduce such  auxiliary
superfields and couplings with them, but up to now all constructed
models have been restricted to have conformally flat sigma models
in the bosonic sector. This restriction has an evident source - it
has been known for a long time that all linear \Nf supermultiplets
can be obtained through a dualization procedure from the \Nf
``root'' supermultiplet  -- \Nf hypermultiplet
\cite{hyper1,hyper2,hyper3,hyper4,root,DI1}, while the bosonic
part of the general hypermultiplet action is always conformally  flat.
The only way to escape this flatness situation is to use
nonlinear supermultiplets \cite{AS,nelin1,NLM2}, instead of linear
ones.

The main aim of the present paper is to construct the Lagrangian formulation of \Nf supersymmetric mechanics
on hyper-K\"{a}hler spaces in the non-Abelian background gauge fields. To achieve this goal we will perform two steps
\begin{itemize}
\item First, we introduce the coupling of the ``matter''  \Nf tensor supermultiplet ${\cal V}^{ij}$ with an
auxiliary fermionic supermultiplet $\Psi^{\halpha}$ in a such way to have  the proper action for matter fields
and isospin variables \p{intro2}. This procedure was developed in \cite{bk}. On this step we still have
conformally flat three-dimensional sigma-model in the bosonic sector.
\item As the next step, following \cite{AS}, we
dualize the auxiliary component $A$ of the tensor supermultiplet into a fourth physical boson, finishing
with the action having  the
hyper-K\"{a}hler sigma-model in the bosonic sector.
\end{itemize}

The resulting action contains a wide class of \Nf supersymmetric
mechanics describing the motion of an isospin-carrying particle
over spaces with non-trivial hyper-K\"{a}hler geometry and in the presence of a
non-Abelian background gauge field. In two examples we discussed
in details, these backgrounds correspond to the field of the BPST
instanton in the flat and Taub-NUT spaces.

\setcounter{equation}0
\section{\Nf supersymmetric isospin particles in conformally flat spaces}
One of the possible ways to incorporate the isospin-like variables
in the Lagrangian of supersymmetric mechanics is to couple the
basic superfields with auxiliary fermionic superfields
$\Psi^\halpha, {\bar \Psi}_\halpha$, which contain these isospin
variables \cite{bk}. Such a coupling, being written in a standard
\Nf superspace, has to be rather special, in order to provide a
kinetic term of the first order in time derivatives for the
isospin variables and to describe the auxiliary fermionic
components present in $\Psi^\halpha, {\bar \Psi}_\halpha$.
Following \cite{bk}, we introduce the coupling of auxiliary $\Psi$
superfields with some arbitrary, for the time being, \Nf
supermultiplet $X$  as
\be\label{actionX}
S_c=-\frac{1}{32}\int dt
d^4\theta \; \left( X +g\right)  \Psi^\halpha {\bar\Psi}_\halpha,
\qquad g=const.
\ee
The $\Psi$ supermultiplet is subjected to the
irreducible conditions \cite{ikl1}
\be\label{Psi} D^i
\Psi^1=0,\quad D^i \Psi^2+\bD{}^i \Psi^1=0, \quad \bD_i \Psi^2=0,
\ee
and thus it contains four fermionic and four bosonic
components
\be\label{psi} \psi^\halpha=\Psi^\halpha|, \qquad u^i
=-D^i {\bar\Psi}{}^2|,\quad {\bar u}_i= \bD_i \Psi^1|,
\ee
where the symbol $|$ denotes the $\theta=\tb=0$ limit and \Nf covariant
derivatives obey standard relations
\be
\left\{ D^i,
\bD_j\right\}=2 \im \delta^i_j
\partial_t.
\ee
It has been demonstrated in \cite{bk} that if the \Nf
superfield $X$ is subjected to the constraints
\cite{ikl1,leva}
\be\label{X} D^i D_i X=0,\quad \bD_i \bD{}^i X=0,\quad \left[
D^i,\bD_i\right] X=0,
\ee
then the component action which follows
from \p{actionX} can be written as
\bea\label{actionXc}
S_c&=&\int dt\left[ -(x+g)\left({ \rho}{}^1{\brho}{}^2-{
\rho}{}^2{\brho}{}^1\right)-\frac{i}{4} (x+g) \left( {\dot u}{}^i
{\bar u}_i-
u^i\dot{\bar u}_i\right)+\frac{1}{4}A_{ij}u^i{\bar u}{}^j \right.\nn\\
&+& \left. \frac{1}{2}\eta_i\left({\bar u}{}^i \brho{}^2+u^i
\rho{}^2\right)+\frac{1}{2}\bar\eta{}^i\left(u_i \rho{}^1+{\bar
u}_i\brho{}^1\right) \right],
\eea
where the new fermionic
components $\rho^\halpha,{\bar\rho}_\halpha$  are defined as
\be\label{rho} \rho^\halpha ={\dot\psi}{}^\halpha, \quad
{\bar\rho}_\halpha={\dot{\bar\psi}}_\halpha.
\ee
The components of the
superfield $X$ entering the action \p{actionXc} have been
introduced as
\be\label{compX}
 x= X|,\quad  A_{ij} = A_{(ij)}= \frac{1}{2}\left[ D_i,\bD_j\right] X|,\quad \eta^i= -iD^i X|,\quad
\bar\eta{}_i= -i\bD_i X|.
\ee
What makes the action \p{actionXc}
interesting is that, despite the non-local definition of the
fermionic components  $\rho^\halpha, {\bar\rho}_\halpha$ \p{rho}, the action is
invariant under the following  \Nf supersymmetry transformations:
\bea\label{n4tr1}
&&\delta \rho^1=-\bar\epsilon{}^i \dot{\bar
u}_i,\quad \delta\rho^2=\epsilon_i\dot{\bar u}{}^i,\quad \delta
u^i=-2i\epsilon^i\brho{}^1+2i\bar\epsilon{}^i\brho{}^2,\quad \delta
{\bar u}_i=-2i\epsilon_i\rho{}^1+2i\bar\epsilon_i\rho{}^2, \nn \\
&& \delta x=-i\epsilon_i\eta^i-i\bar\epsilon{}^i\bar\eta_i,\quad
\delta\eta{}^i=-\bar\epsilon{}^i{\dot x}-i\bar\epsilon{}^j
A^i_j,\quad \delta \bar\eta_i=-\epsilon_i{\dot x}+i\epsilon_j
A_i^j,\quad \delta A_{ij} = -\epsilon_{(i}\dot\eta_{j)} +
\bar\epsilon_{(i}\dot{\bar\eta}{}_{j)}.
\eea
In the action \p{actionXc} the fermionic fields $\rho^\halpha,\brho{}_\halpha$
are auxiliary ones, and thus they can be eliminated  by their
equations of motion
\be\label{ad5a}
\rho^1=\frac{1}{2(x+g)}\eta_i{\bar u}{}^i, \qquad
\rho^2=-\frac{1}{2(x+g)}\bar\eta{}^i{\bar u}_i. \ee
{}Finally, the
action describing the interaction of $\Psi$ and $X$
supermultiplets acquires a very simple form
\be\label{actionXcfin}
S_c=\frac{1}{4}\int dt\left[
 -i (x+g) \left( {\dot u}^i {\bar u}{}_i- u^i\dot{\bar u}{}_i\right)+
 A_{ij}u^i{\bar u}{}^j+
\frac{1}{x+g} \eta_i\bar\eta_j\left( u^i {\bar u}{}^j+ u^j{\bar
u}{}^i\right) \right].
\ee
Thus, in the fermionic superfields
$\Psi$ only the bosonic components $u^i,{\bar u}_i$, entering the
action with a kinetic term linear in time-derivatives, survive.

Comparing our action \p{actionXcfin} with \p{intro2} one may see that the
term describing the interaction with the gauge field is just (one has firstly to rescale isospin
variables to bring their kinetic term to the flat form \p{intro2})
\be\label{aaa}
\frac{1}{g+x}\;A_{ij}{\omega}^i{\bar{\omega}}{}^j, \quad {\omega}^i=\sqrt{g+x}\; u^i.
\ee
It is clear that the explicit form of the gauge field could be read off only after
expressing the (for the time being) ``auxiliary'' components $A_{ij}$ in terms of
physical bosons present in the model.
Thus, in order to be meaningful, the action \p{actionX} has to be
extended by the action for the supermultiplet $X$ itself:
\be\label{Xpsi}
S=S_x+S_c =-\frac{1}{32}\int dt d^4\theta {\cal F}(X)+S_c ,
\ee
where ${\cal F}(X)$ is an arbitrary function depending on superfield $X$.

If the superfield $X$ obeying \p{X} is considered as an independent
superfield, then the components $A_{ij}$ \p{compX} are auxiliary ones, and they
have to be eliminated by their equations of motion. The resulting
action describes \Nf supersymmetric mechanics with one physical boson $x$ and four physical fermions
$\eta^i, {\bar\eta}_j$ interacting with isospin variables $u^i, {\bar u}_i$.
Just this system has been considered in \cite{{FIL},{FIL1},{bk}}.

It is clear that treating the scalar bosonic superfield $X$ as an
independent one is too restrictive, because the constraints \p{X}
leave in this supermultiplet only one physical bosonic component
$x$, which is not enough to describe the isospin particle. In the
present approach, the way to overcome this limitation was proposed
in \cite{{sutulin1},{sutulin2}}. The key point is to treat the
superfield $X$ as a composite one, constructed from \Nf
supermultiplets with a larger number of physical bosons. The
reasonable superfield from which it is possible to construct the
superfield $X$ is the \Nf tensor supermultiplet ${\cal V}^{ij}$
\cite{{v1},{v1a}}.\vspace{0.5cm}

\noindent{\bf Tensor supermultiplet}\\
The \Nf tensor supermultiplet is described by the triplet of
bosonic \Nf superfields ${\cal V}^{ij}={\cal V}^{ij}$ subjected to
the constraints
\be\label{V}
D^{(i}{\cal V}^{jk)}=\bD{}^{(i}{\cal V}^{jk)}=0, \qquad
\left( {\cal V}^{ij}\right)^\dagger = {\cal V}_{ij},
\ee
which leave in ${\cal V}^{ij}$ the following
independent components:
\be\label{v}
v^a=-\frac{\im}{2}\left( \sigma^a\right)_i{}^j{\cal V}_j^i|,\quad
\lambda^i=\frac{1}{3}D^j{\cal V}^i_j|,\quad
\blam{}_i =\frac{1}{3}\bD_j{\cal V}^j_i|, \quad A=\frac{\im}{6}D^i  \bD{}^j
{\cal V}_{ij}|.
\ee
Thus, its off-shell component field content is
$(3,4,1)$, i.e. three physical $v^a$ and one auxiliary $A$ bosons
and four fermions $\lambda^i, \blam{}_i$ \cite{{v1},{v1a}}. Under
\Nf supersymmetry these components transform as follows:
\bea\label{Vtr}
&&
\delta v^a=\im \epsilon^i
(\sigma^a)_i^j\blam_j-\im \lambda^i(\sigma^a)_i^j
\bar\epsilon_j,\quad
\delta A=\bar\epsilon_i {\dot\lambda}{}^i-\epsilon^i\dot\blam_i, \nn \\
&&
\delta\lambda^i=\im \epsilon^i A+\epsilon^j(\sigma^a)_j^i{\dot
v}{}^a,\quad \delta\blam_i=-\im \bar\epsilon_i
A+(\sigma^a)_i^j\bar\epsilon_j{\dot v}{}^a .
\eea
Now one may check that the composite superfield
\be\label{XV}
X=\frac{1}{|\cal
V|} \equiv \frac{1}{\sqrt{{\cal V}^a {\cal V}^a}},
\ee
where ${\cal V}^a=-\frac{\im}{2}\left( \sigma^a\right)_i{}^j{\cal
V}_j^i$, obeys \p{X} in virtue of \p{V}. Clearly, now all
components of the $X$ superfield, i.e. the physical boson $x$,
fermions $\eta^i, \bar\eta_i$ and auxiliary fields $A^{ij}$
\p{compX} are expressed through the components of the ${\cal
V}^{ij}$ supermultiplet \p{v} as
\bea\label{rel1}
&&
x=\frac{1}{|v|}, \qquad \eta^i = \frac{v^a}{|v|^3}
(\lambda\sigma^a)^i, \quad
\bar\eta_i =   \frac{v^a}{|v|^3} (\sigma^a\blam)_i, \nn \\
&&
A^i_j =-3\frac{v^a
v^b}{|v|^5}(\lambda\sigma^a)^i(\sigma^b\blam)_j-
\frac{v^a(\sigma^a)^i_j}{|v|^3}A+\frac{1}{|v|^3}\epsilon^{abc}v^a{\dot
v}{}^b(\sigma^c)^i_j+ \frac{1}{|v|^3}\left( \delta^i_j
\lambda^k\blam_k-\lambda_j\blam^i\right).
\eea
The third term in the above expression for $A^{ij}$ being substituted in the action
\p{actionXcfin} provides us with the term
\be\label{WY}
 {\cal A}_b^a I^a{\dot v}{}^b=\frac{1}{(g+\frac{1}{|v|})|v|^3}\epsilon^{cab}v^c\; I^a \;{\dot
v}{}^b, \qquad I^c= \frac{1}{2}\,{\omega}^i (\sigma^c)_i^j {\bar {\omega}}_j,\; {\omega}{}^i=\sqrt{g+\frac{1}{|v|}}u^i
\ee
Thus, for the $g=0$ case our gauge field describes the magnetic field of  a Wu-Yang monopole \cite{WY}.

Finally, one
should note that, while dealing with the tensor supermultiplet
${\cal V}^{ij}$, one may generalize the $S_x$ action \p{Xpsi} to
have the full action in the form
\be\label{Vpsi}
S=S_v+S_c=-\frac{1}{32}\int dt d^4\theta {\cal F}({\cal V}) +S_c,
\ee
where ${\cal F}({\cal V})$ is now an arbitrary function of
${\cal V}{}^{ij}$. After eliminating the auxiliary component $A$
in the component form of  \p{Vpsi} we will obtain the action
describing the \Nf supersymmetric three-dimensional isospin
particle moving in the magnetic field of a Wu-Yang monopole and in
some specific scalar potential \cite{sutulin1}
\footnote{An alternative description of the same system has been recently
constructed in \cite{Ikon}}.

To close this Section one should mention that, while dealing with
the tensor supermultiplet ${\cal V}^{ij}$, the structure of the action $S_c$
\p{actionX} can be further generalized to be \cite{sutulin2}
\be\label{actionY}
S_c=-\frac{1}{32}\int dt
d^4\theta \; Y \Psi^\halpha {\bar\Psi}_\halpha,
\ee
with $Y$ obeying
\be\label{genY}
\frac{\partial^2}{\partial v_a \; \partial v_a} Y = 0.
\ee
Clearly, our choice $Y=\frac{1}{|v|}+g$   corresponds to spherically-symmetric
solutions of \p{genY}.

\setcounter{equation}0
\section{Hyper-K\"{a}hler sigma model with isospin variables}
One of the most attractive features of our approach is the unified
structure of the action $S_c$ \p{actionX} which has the same form
for any type of supermultiplets, which we are using to construct a
composite superfield $X$. Just this opens the way to relate the
different systems via duality transformations. Indeed, it has been
known for a long time \cite{GR, Tolik, FG, FG1, root, DI1} that in
one dimension one may switch between supermultiplets with a
different number of physical bosons, by expressing the auxiliary
components through the time derivative of physical bosons, and
vice versa. Here we will use this mechanism to obtain the action with four physical
bosonic components starting from the action for the tensor multiplet \p{Vpsi}.  In what follows, to make some
expressions more transparent, we will use, sometimes,  the
following stereographic coordinates for the bosonic components of
tensor supermultiplet \p{v}:
\be
V^{11}=2 \im \frac{e^u}{1+\Lambda\bLam} \Lambda,\qquad V^{22} = -2 \im \frac{e^u}{1+\Lambda\bLam} \bLam,\qquad
 V^{12}=-\im e^u \left( \frac{1-\Lambda\bLam}{1+\Lambda\bLam}\right). \label{stv}
\ee\vspace{0.5cm}

The crucial step in the dualization of the last auxiliary field $A$ \p{rel1} into a fourth
physical boson is  the transformation property of the following combination:
\be\label{fg1}
{\hat A}= B_a {\dot v}{}_a - f_{,a} (\lambda \sigma^a
\bar\lambda)+f A, \qquad f_{,a}\equiv \frac{\partial}{\partial v_a}
\ee
where
\be\label{fH}
f=\frac{1}{|v|}\,,\quad
\mbox{ and },\quad B_1=-\frac{v_2(v_3+|v|)}{(v_1^2+v_2^2)|v|}\,,\quad
B_2=\frac{v_1(v_3+|v|)}{(v_1^2+v_2^2)|v|}\,,\quad B_3=0.
\ee
Indeed, one may easily check that $\hat A$ transforms as a full time derivative under
\Nf supersymmetry. Therefore, in accordance with the general idea of the dualization procedure one
may replace $\hat A$ by a new physical bosonic field $\phi$ as
\be\label{add1}
\hat A = \dot\phi.
\ee
What is much more interesting is that the choice for $f$ and $B_a$ in \p{fH} is not
unique.  It has been proved in \cite{AS} that ${\hat A}$ \p{fg1}
transforms as a full time derivative, if the
functions $f$ and $B_a$ satisfy the equations
\be\label{hkcond}
\triangle_3 f \equiv f_{,aa} = 0, \qquad f_{,a}=\epsilon_{abc}
B_{c,b}.
\ee
Thus, one may construct a more general action for
four-dimensional \Nf supersymmetric mechanics using the component
action for the tensor supermultiplet and substituting there the
new dualized version of the auxiliary component $A$ \p{fg1}.

Integrating over theta's in \p{Vpsi} and eliminating the auxiliary
fermions $\rho^{\halpha}$ \p{ad5a}, \p{rel1}, we will get the
following component action for the tensor supermultiplet:
\bea\label{hk}
S&=&\frac{1}{8} \int dt \left[ F \left( {\dot
v}_a{\dot v}_a +A^2\right) +\im \left( {\dot\xi}{}^i
\bxi_i-\xi^i\dot{\bxi}_i\right)+\im \epsilon_{abc}
\frac{F_{,a}}{F}{\dot v}_b\Sigma_c -
\im \frac{F_{,a}}{F}\Sigma_a A -\frac{1}{6} \frac{\triangle_3 F}{F^2}\Sigma_a\Sigma_a \right. \nn \\
&& -2 \im \left( {\dot w}{}^i{\bar w}_i -w^i{\dot{\bar w}}_i\right) +4 \frac{1+3 g |v|}{F (1+g |v|)^2 |v|^4}
\left(v_a I_a\right)\left(v_b\Sigma_b\right)- 4\frac{g}{F (1+g|v|)^2 |v|}\left( I_a \Sigma_a\right) \nn \\
&& \left. -\frac{4\im}{(1+g|v|)|v|^2}\left(v_a I_a\right)\; A+
\frac{4\im}{(1+g|v|)|v|^2}\epsilon_{abc}v_a{\dot v}_b I_c \right],
\eea where \be\label{defs1} F=\triangle_3\; {\cal F}({\cal V})|,
\qquad I^a=\frac{\im}{2}\left( w \sigma^a  {\bar w}\right), \qquad
\Sigma^a =-\im \left( \xi \sigma^a  {\bar \xi}\right),
\ee
and the re-scaled  fermions and isospin variables are chosen to be
\be
\xi^i = \sqrt{F}\; \lambda^i, \qquad w^i
=\sqrt{g+\frac{1}{|v|}}\;u^i.
\ee
Substituting \p{fg1} into
\p{hk}, we obtain  the resulting action
\bea\label{hkf}
S&=&
\frac{1}{8} \int dt \left[ F \left( {\dot v}_a{\dot v}_a+
\frac{1}{f^2}\left( \dot\phi - B_a {\dot v}_a\right)^2 \right)+\im
\left( {\dot\xi}{}^i \bxi_i-\xi^i\dot{\bxi}_i\right)
-2 \im \left( {\dot w}{}^i{\bar w}_i -w^i \dot{\bar w}_i\right) \right. \nn \\
&& -\im \left[ \frac{1}{f}\delta_{ab}\left(\dot\phi-B_c{\dot v}_c\right)+\epsilon_{abc}{\dot v}_c\right]
\left( \frac{F_{,a}}{F}\Sigma_b +\frac{4}{(1+g |v|)|v|^2} v_a I_b \right) \nn \\
&& +\frac{4}{F} \frac{1+3 g |v|}{(1+g|v|)^2 |v|^4} \left(v_a I_a\right) \left( v_b \Sigma_b\right)-
\frac{1}{F} \frac{4g}{(1+g |v|)^2 |v|}\left( I_a \Sigma_a\right) \nn \\
&& \left. +\frac{1}{3 F^2}\left( \frac{F_{,a}f_{,a}}{f} -\frac{F
f_{,a}f_{,a}}{f^2}- \frac{1}{2} \triangle_3 F \right) \Sigma_b
\Sigma_b \right].
\eea
The action \p{hkf} is our main result. It
describes a motion of a \Nf supersymmetric four-dimensional isospin carrying
particle in the non-Abelian gauge field with potential
\be\label{newA}
{\cal A}_\mu {\dot x}_\mu=  -\im \left[ \frac{1}{f}\delta_{ab}\left(\dot\phi-B_c{\dot v}_c\right)+\epsilon_{abc}{\dot v}_c\right]
\frac{4}{(1+g |v|)|v|^2} v_a I_b ,
\ee
where $x_\mu=(v_a, \phi)$. The four-dimensional bosonic metric of our model is
defined in terms of two functions: the bosonic part of the
pre-potential $F$ \p{defs1} and the harmonic function $f$
\p{hkcond}. The supersymmetric version of the coupling with the
monopole (second line in the action  \p{hkf}) is defined by the
same harmonic function $f$ and the coupling constant $g$. In the
more general case \p{actionY}, we will have two harmonic functions
- $f$ and $Y$, besides the pre-potential $F$.

Among all possible systems with the action \p{hkf} there is a very
interesting sub-class which corresponds to hyper-K\"{a}hler sigma
models in the bosonic sector. This case is distinguished by the
condition
\be\label{HK}
F=f.
\ee
Clearly, in this case the bosonic
kinetic term of the action \p{hkf} acquires the familiar form of
the one dimensional version of the general Hawking-Gibbons
solution for four-dimensional hyper-K\"{a}hler metrics with one
triholomorphic isometry \cite{GH}:
\be\label{GH}
S_{kin}=\frac{1}{8} \int dt \left[ f {\dot v}_a{\dot v}_a+
\frac{1}{f}\left( \dot\phi - B_a {\dot v}_a\right)^2
\right],\qquad \triangle_3 f=0,\; \mbox{rot } \vec{B}
=\vec{\nabla} f.
\ee
It is worth to note that the bosonic part of
\Nf supersymmetric four dimensional sigma models in one dimension
does not necessarily have to be a hyper-K\"{a}hler one. This fact
is reflected in the arbitrariness of the pre-potential $F$ in the
action \p{hkf}. Only under the choice $F=f$ the bosonic kinetic
term is reduced to the Gibbons-Hawking form \p{GH}. Let us note that
for hyper-K\"{a}hler cases the four-fermionic term in the action
\p{hkf} disappears. This fact has been previously established in
\cite{AS}. Now we can see that the additional interaction with
background non-Abelian gauge field does not destroy these nice
properties.

Among all possible bosonic metrics one may easily find the following interesting ones.\vspace{0.5cm} \\
\noindent{\bf Conformally flat spaces.}\\
There are two choices for the function $f$ which correspond to the conformally flat metrics in the bosonic sector. Both of them with
\be\label{1}
f=\frac{1}{|v|}\qquad \mbox{ and } f= const, \qquad B_a = 0
\ee
give rise to supersymmetric mechanics with the hypermultiplet describing the particle
moving in the field of a BPST instanton \cite{sutulin2}.

Let us remind that in both these cases we have not specified the
pre-potential $F$ yet. Therefore, the full metrics in the bosonic
sector is defined up to this function.
\vspace{0.5cm} \\
\noindent{\bf Taub-NUT space.} \\
One should stress that the previous two cases are unique, because
only for these choices of $f$  the resulting action \p{hkf} can be
formulated in terms of linear \Nf hypermultiplet \cite{hyper1,hyper2,hyper3,hyper4}. With
other solutions for $f$ we come to the theory with the nonlinear
\Nf hypermultiplet \cite{AS,nelin1}. Among the possible solutions
for $f$ which belongs to this new situation the simplest one
corresponds to one center Taub-NUT metrics with
\be\label{3}
f=p_1+\frac{p_2}{|v|}, \qquad p_1,p_2=\mbox{const}.
\ee
In order
to achieve the maximally symmetric case, we will chose these
constants as
\be\label{3a}
p_1=g, \quad p_2 =1 \quad \rightarrow
\quad f=g+\frac{1}{|v|}.
\ee
With such a definition $f$ coincides
with the function $Y=g+\frac{1}{|v|}$ \p{actionY} entering in our
basic action $S_c$ in \p{actionX}, \p{XV}. To get the Taub-NUT
metrics, one has also to fix the pre-potential $F$  to be equal to
$f$. The resulting action which describes the \Nf supersymmetric
isospin carrying particle moving in a Taub-NUT space reads
\bea\label{TN}
S_{Taub-NUT}&=& \frac{1}{8} \int dt \left[  \left(
g+\frac{1}{|v|}\right)  {\dot v}_a{\dot v}_a+ \frac{1}{\left(
g+\frac{1}{|v|}\right)}\left( \dot\phi - B_a {\dot v}_a\right)^2
+\im \left( {\dot\xi}{}^i \bxi_i-\xi^i\dot{\bxi}_i\right)
-2 \im \left( {\dot w}{}^i{\bar w}_i -w^i \dot{\bar w}_i\right) \right. \nn \\
&& +\frac{\im}{(1+g |v|)|v|^2} \left[ \frac{v_a}{\left( g+\frac{1}{|v|}\right)}\left(\dot\phi-B_c{\dot v}_c\right)-\epsilon_{abc}v_b{\dot v}_c\right]
\left( \Sigma_a -4 I_a \right) \nn \\
&&\left. +\frac{4(1+3 g |v|)}{(1+g|v|)^3 |v|^3} \left(v_a I_a\right) \left( v_b \Sigma_b\right)-
 \frac{4g}{(1+g |v|)^3}\left( I_a \Sigma_a\right) \right].
\eea The bosonic term in the second line of this action can be
rewritten as \be\label{inst} {\cal A}_a I_a =\frac{\im}{2}\left[
\frac{1}{f}\; \frac{\partial \log f}{\partial v_a}\left( \dot\phi
- B_c {\dot v}_c\right)-\epsilon_{abc} \frac{\partial \log
f}{\partial v_b}{\dot v}_c\right] I_a , \ee where $f$ is defined
in \p{3a}. In this form the vector potential ${\cal A}_a$
coincides with the potential of a Yang-Mills $SU(2)$ instanton in
the Taub-NUT space \cite{inst1,inst2}, if we may view $I_a$, as
defined in \p{defs1}, as proper isospin matrices. The remaining
terms in the second and third lines of \p{TN} provide a \Nf
supersymmetric extension of the instanton.

Finally, to close this Section, let us note that more general
non-Abelian backgrounds can be obtained from the multi-centered
solutions of the equation for the harmonic function $Y$ \p{genY},
which defined the coupling of the tensor supermultiplet with
auxiliary fermionic ones.
Thus, the variety models we
constructed are defined through three functions: pre-potential ${\cal F}$ \p{Vpsi} which is an arbitrary function,
3D harmonic function $Y$ \p{actionY}, \p{genY} defining the coupling with isospin variables and, through,
again 3D harmonic, function $f$ \p{fg1}, \p{hkcond} which appeared during the dualization of the auxiliary
component of the tensor supermultiplet.
 It is clear that
we can always redefine $F$ to be $F={\tilde F}f$. Thus, all our
models are conformal to hyper-K\"{a}hler sigma models with \Nf
supersymmetry describing the motion of a particle in the
background of non-Abelian field of the corresponding instantons.

\setcounter{equation}0
\section{Conclusion}

In the present paper we constructed the Lagrangian formulation of
\Nf supersymmetric mechanics with hyper-K\"{a}hler sigma models in
the bosonic sector in the non-Abelian background gauge field. The
resulting action includes the wide class of \Nf supersymmetric
mechanics describing the motion of an isospin-carrying particle
over spaces with non-trivial geometry. In two examples we
discussed in details, the background fields are identified with
the field of BPST instantons in the flat and Taub-NUT spaces.

The approach we used in the paper utilized two ideas: (i) the
coupling of matter supermultiplets with an auxiliary fermionic
supermultiplet $\Psi^{\halpha}$ containing on-shell four physical
fermions and four auxiliary bosons playing the role of isospin
variables and (ii) the dualization of the auxiliary component $A$
of the tensor supermultiplet into a fourth physical boson. The
final action we constructed contains three arbitrary functions:
the pre-potential ${\cal F}$, a 3D harmonic function $Y$ which
defines the coupling with isospin variables and, again 3D
harmonic, a function $f$ which appeared during the dualization of
the auxiliary component of the tensor supermultiplet. The
usefulness of the proposed approach is demonstrated by the
explicit example of the simplest system with non-trivial geometry
- the \Nf supersymmetric action for one-center Taub-NUT metrics.
We identified the background gauge field in this case, which
appears automatically in our framework, with the field of the BPST
instanton in the Taub-NUT space. Thus, one may hope that the other
actions will possess the same structure.

Of course, the presented results are just preliminary in the full
understanding of \Nf supersymmetric hyper-K\"{a}hler sigma models
in non-Abelian backgrounds. Among interesting, still unanswered
questions, one should note the following ones:
\begin{itemize}
\item The full analysis of the general coupling with an arbitrary harmonic function $Y$ has yet to be
carried out.
\item The structure of the background gauge field has to be further clarified: is this really the field of some
monopole (instanton) for any hyper-K\"{a}hler metrics?
\item The Hamiltonian construction is really needed. Let us note that the Supercharges have to be very
specific, because the four-fermions coupling is absent in the case
of HK metrics!
\item  It is quite interesting to check the existence of  the conserved Runge-Lenz vector in the
fully supersymmetric version.
\item The explicit examples of other hyper-K\"{a}hler metrics (say, multi-centered Eguchi-Hanson and Taub-NUT ones) would be
very useful.
\item The questions of quantization and analysis of the spectra, at least in the cases of well known, simplest
hyper-K\"{a}hler metrics, are doubtless urgent tasks.
\end{itemize}

Finally, let us stress that our construction is restricted to the
case of hyper-K\"{a}hler metrics with one translational
(triholomorphic) isometry. It will be very interesting to find a similar
construction applicable to the case of geometries with rotational
isometry. We hope this can be done within the approach discussed
in \cite{tri}.

 \setcounter{equation}0
\section*{Acknowledgements}
We thank Andrey Shcherbakov for useful discussions.
S.K. and A.S. are grateful to the Laboratori Nazionali di Frascati for hospitality.
This work was partially supported by the grants RFBF-09-02-01209 and 09-02-91349, by
Volkswagen Foundation grant~I/84 496 as well as by the ERC Advanced
Grant no. 226455, \textit{``Supersymmetry, Quantum Gravity and Gauge Fields''%
} (\textit{SUPERFIELDS}).

\end{document}